\documentstyle[aaspp4,flushrt]{article}
\input epsf.sty

\begin{document}
\title{Effects of Distance Uncertainties on Determinations of 
the Galaxy Peculiar Velocity Field}

\author{Nelson Padilla, Manuel Merch\'an \altaffilmark{1} and 
Diego G. Lambas.\altaffilmark{2}}
\affil{Grupo de Investigaciones en Astronomia Teorica y Experimental,
IATE, Observatorio Astronomico, Laprida 854, 5000 Cordoba Argentina
}

\altaffiltext{1}{CONICOR, C\'ordoba, Argentina}
\altaffiltext{2}{CONICET, Argentina}
\begin{abstract}
Galaxy distance indicators are subject to different uncertainties
and biases which may reflect in the peculiar velocity field. 
In this work, we present different statistical analysis 
applied to peculiar velocities of samples of galaxies 
and mock catalogs taken from numerical simulations 
considering these observational effects.
Our statistical studies comprise the mean relative
velocity and velocity correlation for pairs of galaxies as a function of 
separation, and a bulk flow analysis determined on spheres of $10 h^{-1}$
$Mpc$ radius. 
In order to test the statistical analysis
we use COBE normalized CDM numerical simulations with 
different density parameters and cosmological constant and  
take into account the Tully-Fisher (TF) scatter and a possible TF zero-point
offset, as well as variations in the results of the simulations
arising from different observer positions.
We compare the results of the mock catalogs with samples of
spiral galaxies taken from the Mark III catalog. 
The results of our analysis indicate
the importance of errors in deriving the density parameter $\Omega$ from 
statistics applied to the observational data
and it is found that large values $\Omega \geq 0.8$  
may be derived from the analysis if errors are not taken into account.
However, the observed value of the TF scatter ($\simeq 0.35$ $mag$)    
make CDM models with $\Omega > 0.6$,
inconsistent with the statistical tests involving relative velocities.  
A suitable agreement is found for models with $\Omega \leq 0.6$, 
although requiring a TF 
zero-point offset of the order of a tenth of a magnitude to provide
consistency with the observed flow coherence.
\end{abstract}


\section{Introduction}

Measurements of peculiar velocities provide direct probes of the 
mass distribution in the Universe putting constraints on 
models of large-scale structure formation (e.g. \cite{peebles}, 
\cite{vittorio}).
Advances on new distance indicators
(\cite{djor}; \cite{dresslera}) have allowed estimates
of peculiar flows in the local Universe up to $\sim 50-100 h^{-1}$Mpc (see 
\cite{giova}, or \cite{strauss} for a review).   
The results from using elliptical galaxies
(\cite{dresslerb}) suggest a large coherence length and 
amplitude of the local peculiar velocity 
field. The results from analysis of 1,355 spirals using the Tully-Fisher
relation to determine distances (\cite{mathew}) would lead to an 
even larger coherence length and
a similar amplitude. Other samples are also in agreement with the
Great Attractor findings (\cite{willick}). In a more controversial finding 
\cite{postman} find from analysis of clusters of galaxies that the peculiar 
flows can be coherent over a
much larger scale than the Great Attractor findings.
All the results indicate significant deviations from the Hubble flow 
with a large coherence length.

Since peculiar velocities probe the mass distribution and also depend 
on the density parameter 
$\Omega$ one can determine the latter by comparing the mass distribution 
implied by the velocity field with the observed distribution of galaxies.
The density parameter is thus determined to within the uncertainty of the 
bias factor $b$ by determining the factor $\Omega^{0.6}/b$. 
Other analysis of the peculiar velocity information may be applied
in order to obtain the parameter $\Omega$. 
\cite{berta} developed the POTENT method whereby the mass distribution is 
reconstructed by using the analog of the Bernoulli equation for 
irrotational flows.
They used the method to analyze to great detail the peculiar velocity field
out to about 60$h^{-1}$Mpc (\cite{bertb}). \cite{dekel} 
compared the previously determined velocity field with the 
observed distribution of galaxies and concluded that the best fit 
is with $\Omega^{0.6}/b \simeq 1$.
A similar analysis can be applied to the components of the velocity tensor 
$U_{ij}$ (\cite{gorskia}; \cite{groth}). 
Recently, \cite{kashlinsky} has analyzed this tensor
for the MarkIII peculiar velocity data to reconstruct the large-scale 
power spectrum obtaining consistency within the CDM model with 
$\sigma_{8} \Omega^{0.6} \simeq 0.8$. In a similar study \cite{zaroubi}
obtain a different result $\sigma_{8} \Omega^{0.6} \simeq 0.35$ indicating
the need of further analysis.

In this paper we study the peculiar velocity field  
through a statistical analysis of COBE normalized CDM numerical simulations
and observational data taken from the Mark III catalog.
Our comparison of models and 
observations take into account observational uncertainties and possible 
biases, and
variations of the results according to different observers in fully
non-linear numerical simulations.

\section {Data}

We use the Mark III catalog (\cite{markIII1};\cite{markIII2};\cite{markIII3})
as a suitable data set 
to analyze the peculiar velocity flow.  
This Catalog lists Tully-Fisher and $Dn-\sigma$ distances and radial velocities 
for spiral, irregular, and elliptical
galaxies.
In our analysis we have used only spiral galaxies given their
large number and smooth spatial distribution 
(see Table 1). 

\begin{deluxetable}{cccc}
\tablewidth{35pc}
\tablecaption{Observations: The Mark III spirals}
\tablehead{
\colhead{Subsample}&
\colhead{$N^o$ of Gx.}&
\colhead{TF relation}&
\colhead{$\sigma_{TF}$}
 }
\startdata
 \phm{II} Aaronson et al. Field & 359 &  $M_H=-5.95+10.29 \eta$ & $0.47$ \nl
 \phm{II} Mathewson et al. (1992) & 1355 & $M_I=-5.79+6.8 \eta$ & $0.43$ \nl
 \phm{II} Willick, Perseus Pisces (1991) & 383 & $M_r=-4.28+7.12 \eta$
& 0.38 \nl
 \phm{II} Willick, Cluster Galaxy (1991) & 156 & $M_r=-4.18+7.73 
\eta$ & $0.38$ \nl
 \phm{II} Courteau-Faber (1993) & 326 & $M_r=-4.22+7.73 \eta$ & $0.38$\nl
 \phm{II} Han-Mould et al., Cl. Gx. (1992) & 433 & $M_I=-5.48+7.87 
\eta$ & 0.4 \nl
\enddata
\end{deluxetable}

The velocity parameter $\eta = Log \Delta V - 2.5$ is 
determined either from HI profiles 
or from optical $H_{\alpha}$ rotation curves.
The TF relations and their corresponding scatters 
for the different samples of spiral galaxies
are given by \cite{markIII3} and are shown in Table 1, where the absolute
magnitud $M$ satisfies $M=m-5\log cz$.
The galaxy apparent magnitudes $m$ of the
Tully-Fisher distances are corrected for Galactic
extinction, inclination and redshift
(see Willick et al. 1997 for details).

The selection bias in the calibration of the forward TF relation
can be corrected once the selection function is known.  
But then,
the TF inferred distances and the mean peculiar velocities 
suffer from Malmquist bias.
Suitable  procedures to consider these biases, induced both by
inhomogeneities and selection function, 
have been discussed (see for instance
/cite{freud95} and references therein) where the spatial distribution,
selection effects and observational uncertainties are realistically modeled
through Monte-Carlo simulations.
We have used in our analysis forward TF distances, fully corrected
for Inhomogeneous Malmquist Bias (\cite{markIII1}, \cite{markIII2}, 
\cite{markIII3}).
Nevertheless we have found that the results of the statistics studied in
this work do not change significantly if inverse TF distances are used as
it will be discussed below.

Radial velocities used to infer the peculiar velocity of the galaxies
are referred to
the Cosmic Microwave Background frame.  
It should be remarked that galaxy distance estimates are subject to errors
due to the scatter in the TF relation (\cite{mo}; \cite{willickth}; 
\cite{mathew}) and uncertainties of the TF zero-point (\cite{shanks}; 
\cite{willickth}). Also, the possible presence of
a small fraction of spurious velocities in the data induced by either
galaxy peculiarities or observational errors in distance estimates
(\cite{jacoby}) should be taken into account.

\section{Numerical Simulations}

The N-body numerical simulations  were performed using the Adaptative
Particle-Particle Particle-Mesh (AP3M) code
developed by \cite{couchman}. Initial 
conditions were generated using the Zeldovich approximation and 
correspond to the adiabatic CDM power spectrum 
with different values of $\Omega$ and $\Omega_{\Lambda}$. We have adopted the
analytic fit to the CDM power spectrum given by \cite{sugiyama}:

\begin{equation}
P(k) \propto \frac{k}{[+3.89q+(16.1q)^2+(5.46q)^3+(6.71q)^4]^{1/2}}
\left( \frac{ln(1+2.34q)}{2.34q} \right) ^{2}
\end{equation}

\noindent where $q=\frac{k}{\Gamma h}$Mpc, $\Gamma=\frac{k\theta^2} 
{h \ exp(-\Omega_B- \sqrt{h/0.5} \Omega_B / \Omega)}$, $\theta$ is the
microwave background radiation temperature in units of $2.7K$,
and $\Omega_B=0.0125h^{-2}$ is the value of the baryon density
parameter given by nucleosynthesis.  The normalization of the CDM power
spectrum is imposed by COBE measurements using values of $\Omega$,
$\sigma_{8}$ 
(defined as the mass fluctuations in spheres of radius$=8Mpc$ $h^{-1}$) 
and $h$ 
(the Hubble constant in units of $100$ $km/s/Mpc$)
chosen from Table 1 of \cite{gorskic} and
corresponding to 
ages of the universe $t_0 \simeq$ 12 and 14 Gyr for the $\Lambda=0$ and
$\Lambda \neq 0$ models respectively.
The computational volume is a periodic cube of side length 
300 Mpc.
We have followed the evolution of $N=5\times 10^{5}$ particles with 64
grid per side and a maximum level of refinements of 4. The resulting mass 
per particle is $4.11 \times 10^{12}\Omega_{0} h^{-1} M_{\sun} $. 
The initial conditions correspond to redshift $z=10$ and the evolution
was followed using 1000 steps. At the final step ($z=0$) the linear
extrapolated value of $\sigma_{8}$ is compatible with the normalization
imposed by observed temperature fluctuations in the cosmic background.
In Table 2 are resumed the principal characteristics of the different numerical
simulations.

\begin{deluxetable}{ccccc}
\tablewidth{24pc}
\tablecaption{Simulations}
\tablehead{
\colhead{$\Omega$                }&
\colhead{$\Omega_{\Lambda}$      }&
\colhead{$h_{100}$               }&
\colhead{$\sigma_{8}$            }&
 }
\startdata
  0.3 & 0.7 & 0.65 & 1.05  \nl
  0.4 & 0.0 & 0.65 & 0.75  \nl
  0.5 & 0.0 & 0.60 & 0.90  \nl
  0.6 & 0.0 & 0.60 & 1.12  \nl
  0.7 & 0.0 & 0.60 & 1.32  \nl
  0.8 & 0.0 & 0.55 & 1.325 \nl
\enddata
\end{deluxetable}

\section{Statistical Analysis Outline}

In this section we outline the main statistical procedures followed to analyze 
the galaxy peculiar velocity field, as well as the comparison of model results
and the observations.

Our statistics are based on a direct comparison between the 
observations and mock catalogs derived from the numerical simulations.
The COBE normalized CDM models studied require no strong 
bias, so we have adopted $b=(\delta M/M)/(\delta N/N)=1$ in the simulations.  
This assumption is not crucial in the results of our studies
since we have analyzed mock catalogs with different bias parameters
$.75 <b< 1.35$ finding no significant differences in the statistical results
presented in this work.

Also, we have studied the effects produced by different 
limiting radial distances in our 
comparison of  models and observations by setting two different catalog
depth limits  $d_{lim}=5000$, and $10000$ $km/s$ 
(1764 and 2854 galaxies respectively).

We have considered an analysis of bulk motions as measured by the 
distribution of peculiar velocities of neighbor galaxies
using mean values, dispersions and Kurtosis (as a suitable parameter
to measure departures from gaussianity) to describe the velocity distribution.  
We define shells of radius $\Delta r<2000$ $km/s$ and width $\delta r=200$ 
$Km/s$ centered in each of the galaxies 
of the sample and we compute the  difference
between each galaxy radial velocity in the shell ($v_b$) and the mean motion 
of the galaxies in a sphere of radius $10 \ h^{-1} \ Mpc$ ($\bar{v}_a$). 
We compute the dispersion and kurtosis (D, K respectively) of these 
relative velocities across the shells as a function of $\Delta r$. 

\begin{equation}
D=\sqrt{<(v_b-\bar{v}_a)^2>} \label{d}
\end{equation}
\begin{equation}
K=\frac{(v_b-\bar{v}_a)^4}{D^2}-3    \label{k}
\end{equation}

We have also studied the distribution of the radial velocity moduli of 
the shells in order to derive a quantity related to the coherence of the flow.
We have computed the ratio (M) of the mean of this distribution to the 
dispersion D described in the preceding paragraph.  
The parameter M provides a useful  
information on the coherence of the velocity flow by measuring
the mean systematic to random motion ratio of shells of radius $\Delta r$
(Note that M
differs from the definition of the Cosmic Mach Number given by \cite{ostriker}).

\begin{equation}
M=\frac{<|v_b|>}{D} \label{m}
\end{equation}

We have also provided a different statistical characterization of
the peculiar velocity field through an analysis of 
the distribution of relative radial velocities of pairs of galaxies 
($v_b$ and $v_a$) as a function of separation  (for simplicity
we use the same variable $\Delta r$ as the shell radii of the previously
defined statistics $D$, $K$ and $M$).
We have computed the corresponding dispersion (D1) and kurtosis (K1) and 
we calculate the average modulus of
the pair peculiar velocity difference ($\Delta V$). 
A measurement of the pairwise coherence of the flow is also provided by 
estimating the average of the product of radial peculiar velocities  
of pairs of galaxies with separation $\Delta r$ ($\Pi$).  

\begin{equation}
D1=\sqrt{<(v_b-v_a)^2>} \label{d1}
\end{equation}
\begin{equation}
K1=\frac{(v_b-v_a)^4}{D^2}-3  \label{k1}
\end{equation}
\begin{equation}
\Delta V=<|v_a-v_b|>             \label{del}
\end{equation}
\begin{equation}
\Pi=\sqrt{<v_a v_b>}          \label{pi}
\end{equation}

The D1 parameter corresponds to the mean pairwise velocity dispersion 
usually estimated from the distortion of the correlation function in 
redshift-space.  It should be noted, however, that the D1 
results obtained from the 
peculiar velocity data would not be directly comparable with the 
correlation function distortion findings.  This is mainly
due to the presence of biases and uncertainties 
in the peculiar velocity data and the different hypotheses adopted
in the redshift-space correlation analysis (see for instance \cite{jing}).

We have considered 100 random observers in each numerical simulation by defining
cones with different positions and orientations in our computational volume.  
Also, given the presence of a strong radial gradient in the data
we have attempted to study its possible effects on the determinations of
velocity statistics.  
The observed radial gradient corresponds approximately to a selection
bias due to a magnitude limit cutoff.  This can be seen from the observed
distribution of absolute magnitudes which is nearly gaussian with
mean $\simeq M^*$ (the knee of the Luminosity Function)
and $\sigma \simeq 1.5mag$.  
Nevertheless, for our statistical purposes
it is not necessary to adopt a Monte-Carlo model using the Galaxy
Luminosity Function in the simulations.  This is mainly due to the lack of
strong dependence of observed galaxy luminosities with environment.
Therefore, it suffices to reproduce the observed radial gradient in the
numerical models through a Monte-Carlo rejection algorithm.
Accordingly, we use this procedure to reproduce 
in the mock catalogs derived from the numerical simulations,
the observed galaxy radial gradient.  Furthermore, we restrict 
the resulting number of particles of the 
mock catalogs to be of the order of the number of 
galaxies in the observational sample.

We find that the radial gradient
is very important since it significantly changes the results of the statistics
when errors are included.

In order to quantify the effects of observational errors
in galaxy distance estimates 
we have considered gaussian errors in the galaxy absolute magnitudes
of the TF relation with dispersion  $\Delta M =0.35$ $mag$ 
corresponding to relative errors 
in distances  $\simeq 17 \%$ (\cite{mathew}; \cite{willickth}).  
Namely, we assign to each particle in the mock catalog a new distance
$d_{new}=d*(1+s)$ where $s$ is taken from
a gaussian distribution with dispersion corresponding to the TF uncertainty.  
Then, as the particle peculiar velocities are inferred from the
galaxy redshift and distance, $v_{new}=v-d*s$.

We have also explored the effects produced by a possible TF zero-point offset
(either positive or negative) in the statistical results. 
In our analysis we have considered absolute
magnitude offsets $| \Delta M|=0.15 \ mag$ corresponding to twice the 
$rms$ uncertainty derived in \cite{willickth},
and have been adopted to test their effects on the statistics.

The mean galaxy peculiar velocities are enhanced
due to the scatter of the TF relation by a factor approximately proportional to
their distances.  Since our statistics deal with small angular separations
between the galaxies,
the TF scatter will mostly affect the results involving 
relative radial peculiar velocities.
On the other hand, the coherence of the flow would be significantly enhanced by
an offset in the TF zero-point (either positive or negative) due to the
systematic motion of the galaxies added to the true velocity field.
We have also considered the effects of 
a fraction of spuriously high peculiar velocities in the
Mark III catalog.
To constraint the number of these systematically large velocities in the data
we have added to the peculiar velocities of a given fraction of particles
of the mock catalogs a velocity taken from an even probability distribution 
$p=constant$ within $3\sigma$ and $6\sigma$, and zero elsewhere ($\sigma$ is
the dispersion of the observed distribution of peculiar velocities).
The kurtosis parameter
involves the 4-th power of the difference 
between the observed value of a given quantity and the mean of its
distribution (see eq. \ref{k} or eq. \ref{k1}) 
providing a useful test for the presence of exceedingly large velocities.  
>From a comparison of
observation and model kurtosis values we may pose suitable constrains to 
the fraction of spurious peculiar velocities in the observational data.

\section{Results} 

We have computed the different statistics $K$, $K1$, $D$,
$D1$, $\Delta V$, $\Pi$, and $M$ in both the numerical models and the
observational samples of spirals from Mark III for 8 radial bins. In order
to avoid numerical resolution problems in the simulations 
we analyze relative separations $\Delta r>400$ $km/s$.

In figure 1 we show the kurtosis test results $K$ (eq. \ref{k}) and $K1$ 
(eq. \ref{k1}) for the observations (solid lines) and the models.  

\begin{figure}[bt]
\centerline{\epsfxsize=\textwidth \epsffile{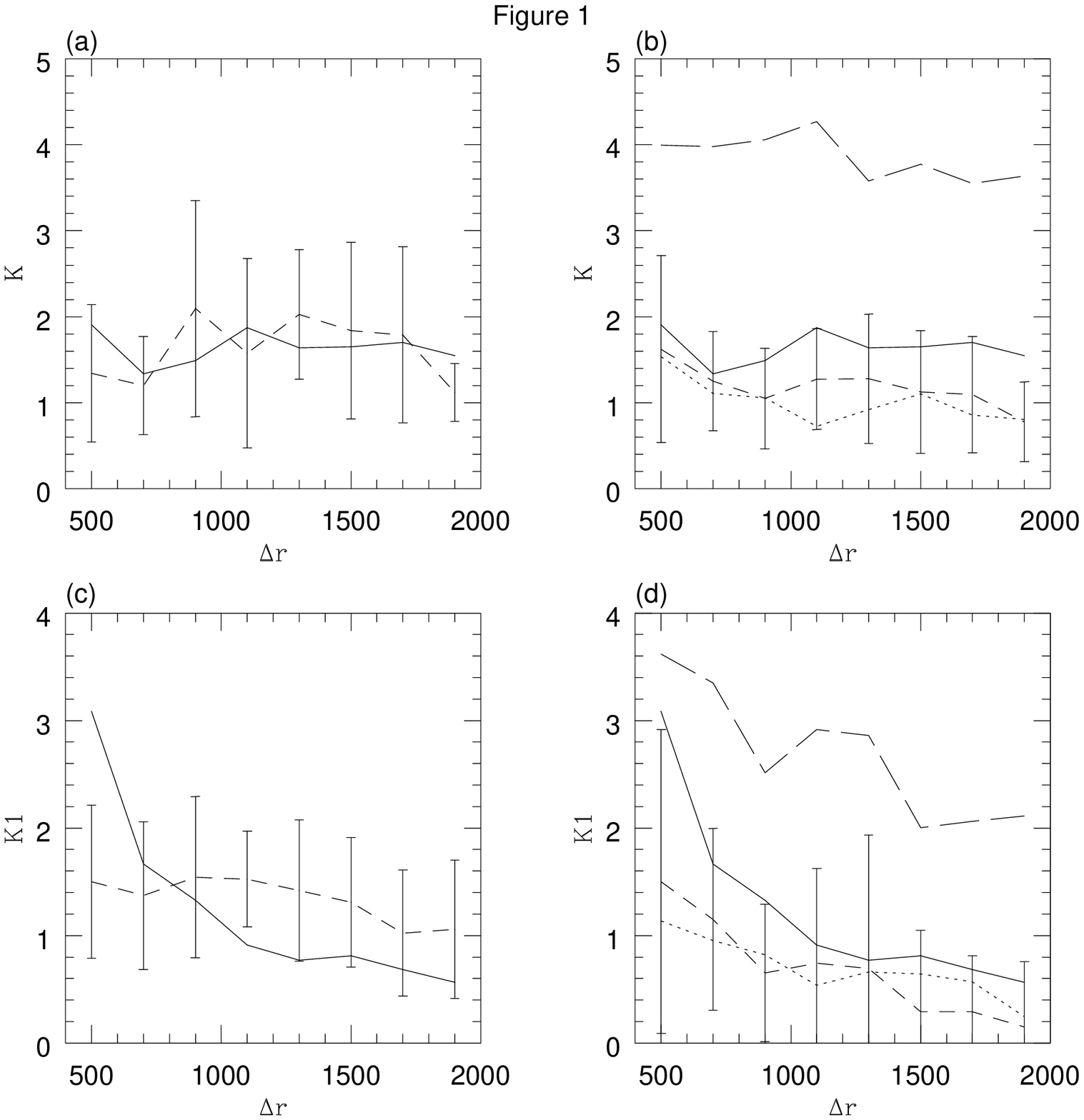}}
\caption{
Kurtosis tests $K$ and $K1$.  Solid lines show the observed values.  
Left panels display the
results with no errors included in the $\Omega=0.4$ CDM model.  Right panels:
dotted lines show the effects of a TF scatter, dashed lines 
the addition of a $+0.15 mag$ TF zero-point offset.  The
long dashed lines show the expected $K$ and $K1$ values under 
the assumption that
$2 \%$ of the galaxies have spurious peculiar velocities (3 to 6 $\sigma$ 
from the mean) due to wrong distance determinations.
Error bars indicate the rms variation arising from different observer 
positions in the $\Omega=0.4$ model.
To avoid confusion in the figures we display only the cases where no errors 
are considered and where a TF scatter with additional positive TF zero-point 
offset is taken into account.}
\label{fig:f1}
\end{figure}

The results for the open $\Omega=0.4$ model without considering peculiar 
velocity errors are displayed with dashed lines in figure 1a and 1c
while the effects of the TF scatter and zero-point offset may be 
appreciated in figures 1b and 1d (dotted and dashed lines
respectively) as well as the effects produced by 
a percentage of exceedingly large
peculiar velocities $v=3\sigma$ to $6\sigma$ (long-dashed lines).
It can be seen in this figure the lack of importance of
the TF scatter and zero point offset 
($0.35$ $mag$ and $+0.15$ $mag$ respectively) in contrast 
to the strong effect of spurious large velocities. A modest fraction
($\simeq 2\%$) of such velocities produce too large values of $K$ and $K1$
in the models compared to observations which suggest this fraction 
as an upper limit for the presence of strongly biased distance estimates.
Also we find similar values of $K$ and $K1$ for the different models 
indicating that kurtosis is not suitable to provide restrictions to 
cosmological parameters. 

In figure 2 we display $D$ (eq. \ref{d}) for the observations (solid lines)
and the $\Omega=0.4$ and $\Omega=0.8$ CDM models.
In figure 2a) no peculiar velocity errors are included
(dashed lines $\Omega =0.4$ model, long dashed lines $\Omega =0.8$ model).
In figure 2b) we show the effects of the TF scatter and zero-point offset. 
The dotted lines show the results of a
$\simeq 17 \%$ relative distance error in the $\Omega = 0.4$ 
model as expected by 
the TF intrinsic scatter ($0.35$ $mag$).    
Additional TF zero-point offsets $\Delta M_{0}= +0.15$ and $-0.15$ $mag$ 
do not significantly change the results and are shown as open and filled
triangles respectively for the $\Omega =0.4$ model. 
Long dashed lines correspond to $\Omega =0.8$ model with TF scatter included
(no TF zero-point offset).
Error bars indicate the rms variation of the results from different 
mock catalog of the models.
It can be seen in this figure that the TF scatter 
strongly dominates this statistical test
allowing low values of $\Omega$ to fit the observations. 

\begin{figure}[bt]
\centerline{\epsfxsize=\textwidth \epsffile{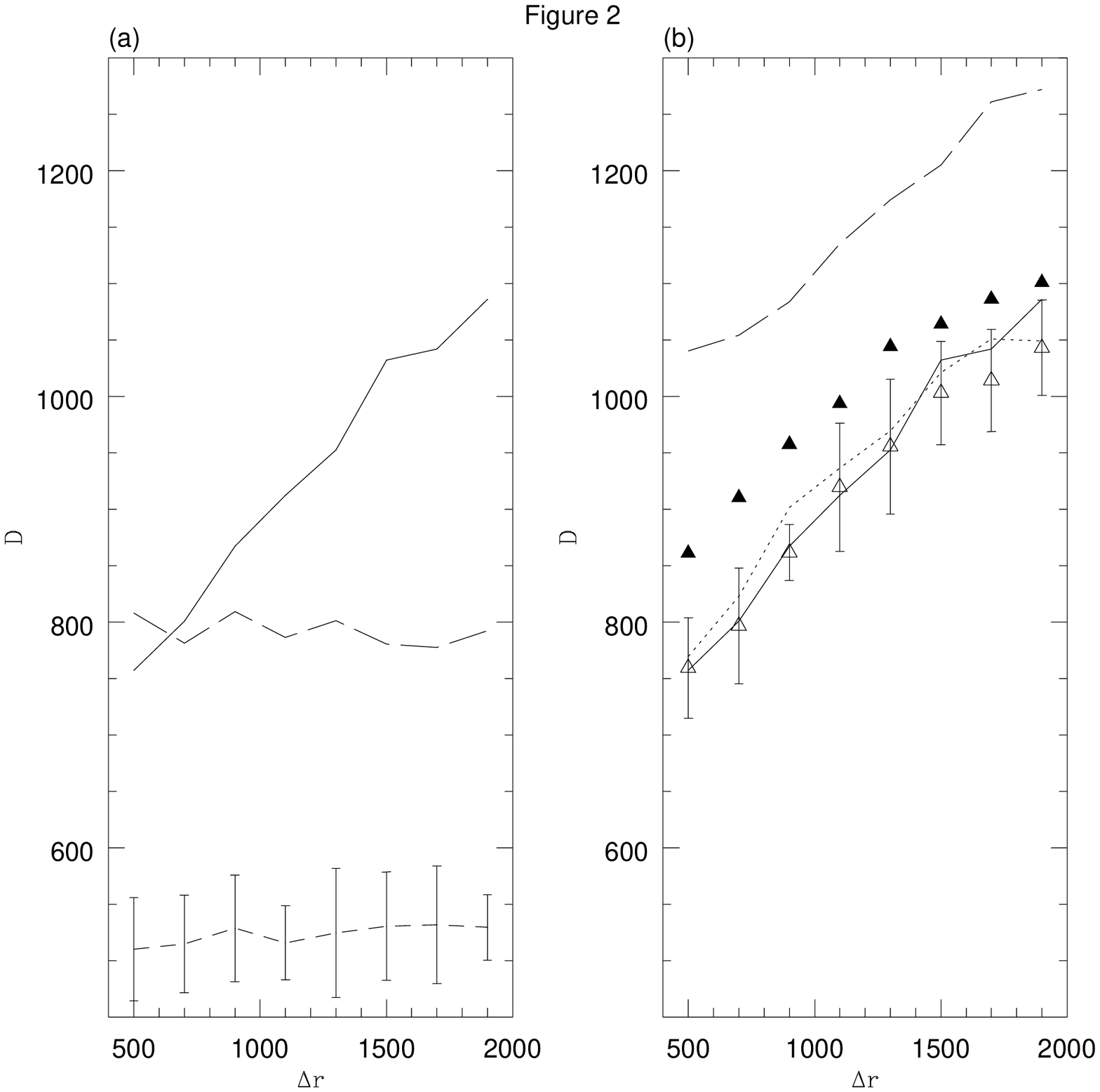}}
\caption{
D test results.  Solid lines show the observed values.
Figure 2a: no errors included in the $\Omega=0.4$ (dashed line) and 
$\Omega=0.8$ (long-dashed line) CDM models.
Figure 2b: effects of a $0.35$ $mag$ TF scatter (dotted line) and additional
$+0.15$ $mag$ and $-0.15$ $mag$ TF zero-point 
offsets (open and filled triangles respectively) in the $\Omega=0.4$ CDM
model.   The long dashed line corresponds to the $\Omega=0.8$ CDM model
with same TF scatter. 
Error bars represent the same as in figure 1.
}
\label{fig:f2}
\end{figure}

Similarly, in figure 3 we show the results of the D1 (eq. \ref{d1}) test in 
the observational data (solid lines) and the corresponding
results of the $\Omega=0.4$ and $\Omega=0.8$ CDM models.
Observational errors are not included in figure 3a and
TF scatter ($\Omega=0.4 $ model, dotted lines)
plus additional $\pm 0.15$ $mag$ TF zero-point offsets 
in the $\Omega=0.4$ model (open and filled triangles).
The results for the $\Omega=0.8$ model with TF scatter are displayed with 
long dashed lines.

\begin{figure}[bt]
\centerline{\epsfxsize=\textwidth \epsffile{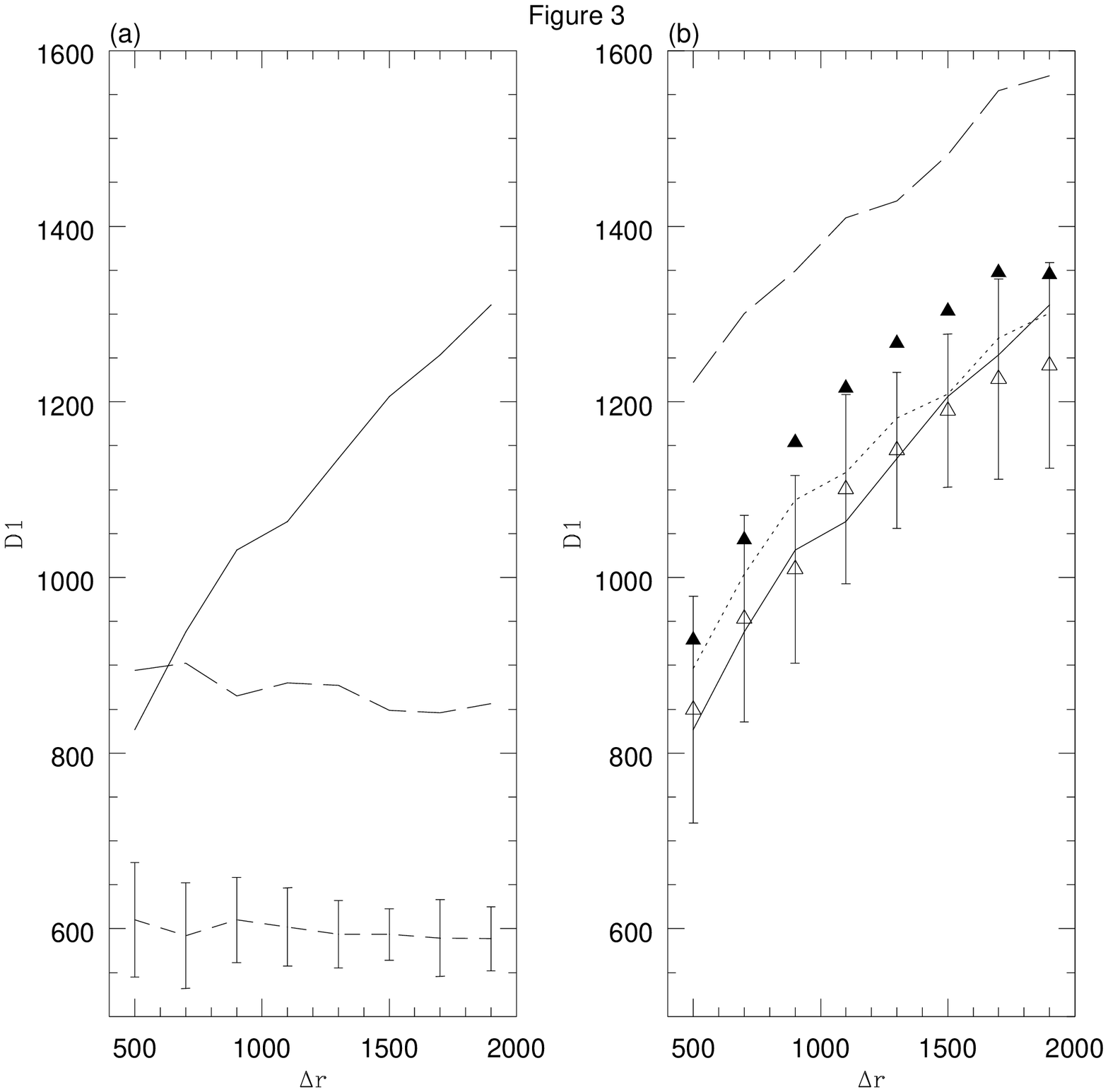}}
\caption{
D1 test results.  Solid lines show the observed values.
Figure 3a: no errors included in the $\Omega=0.4$ (dashed line) and 
$\Omega=0.8$ (long-dashed line) CDM models.
Figure 3b: effects of a $0.35$ $mag$ TF scatter (dotted line) and additional
$+0.15$ $mag$  and $-0.15$ $mag$ TF zero-point offsets 
(open and filled triangles respectively) in the $\Omega=0.4$ CDM
model.  The long dashed line corresponds to the $\Omega=0.8$ CDM model 
with same TF scatter.
Error bars represent the same as in previous figures.
}
\label{fig:f3}
\end{figure}

We show the results of the $\Delta V$ test (eq. \ref{del}) 
for the $\Omega=0.4$ and $\Omega=0.8$ 
models in figure 4 (dotted and dashed lines respectively).
In Figure 4a no errors are included in the models, and
in figure 4b we have included $0.35$ $mag$ TF scatter and 
$\pm 0.15$ $mag$ zero-point offset
with same symbols as in previous figures.

\begin{figure}[bt]
\centerline{\epsfxsize=\textwidth \epsffile{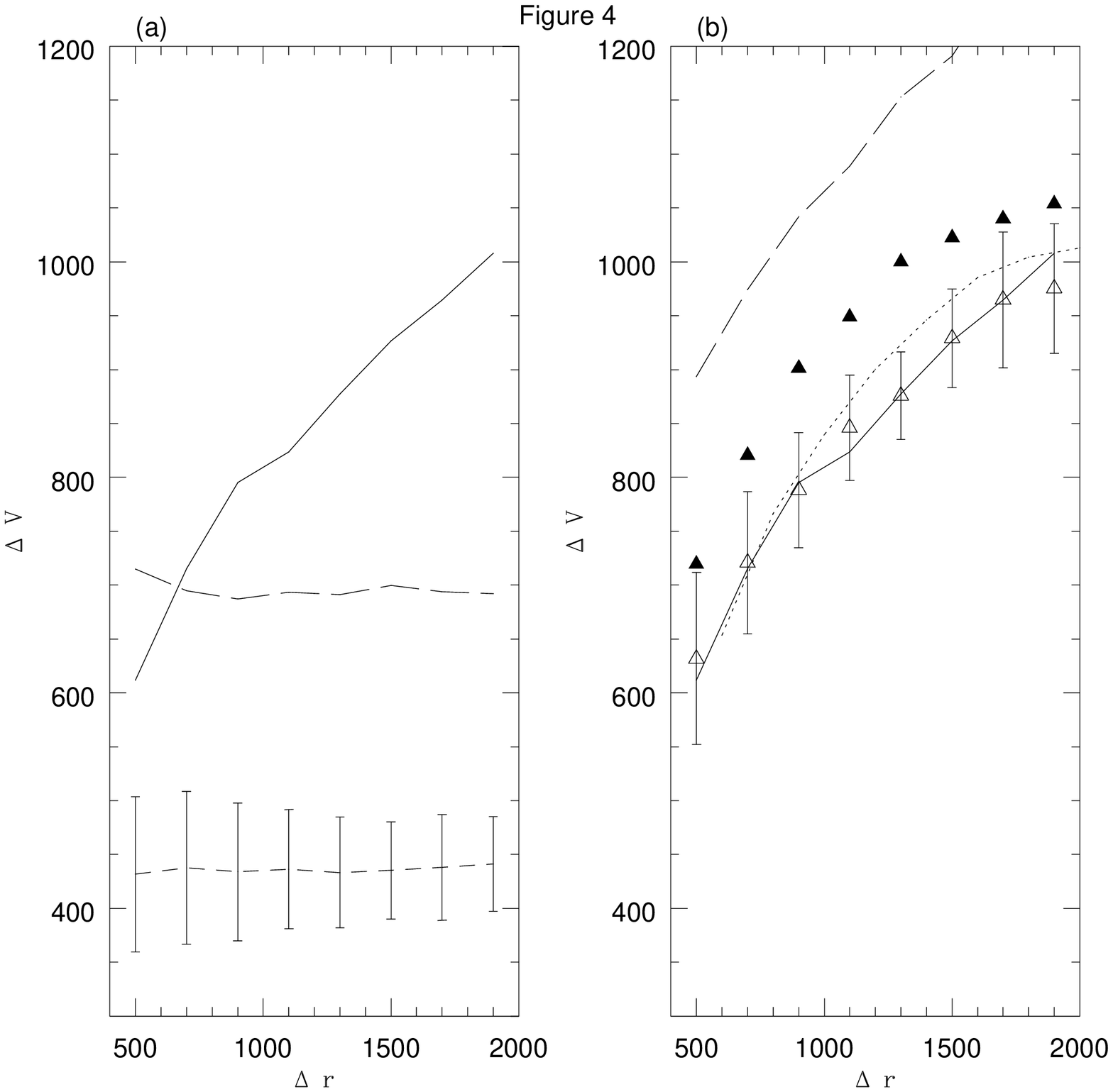}}
\caption{
$\Delta V$ test results.  Solid lines show the observed values.
Figure 4a: no errors included in the $\Omega=0.4$ (dashed line) and 
$\Omega=0.8$ (long-dashed line) CDM models.
Figure 4b: effects of a $0.35$ $mag$ TF scatter (dotted line) and additional
$+0.15$ $mag$ and $-0.15$ $mag$ TF zero-point offsets 
(open and filled triangles respectively) in the $\Omega=0.4$ CDM
model.  The long dashed line corresponds to the $\Omega=0.8$ CDM model
with same TF scatter.
Error bars represent the same as in previous figures.
}
\label{fig:f4}
\end{figure}

It can be appreciated by inspection to figures 2 to 4
the difficulties of matching the observations with
a high density CDM model as well as the adequate fit provided by
the $\Omega=0.4$ model (the flat $\Omega=0.3$ model works very good as well).
It is a remarkable fact that the magnitude and 
shape of the observed $D$, $D1$ and $\Delta V$ as a
function of separation can be adequately reproduced in a low density CDM model
when the TF scatter is taken into account.

In figure 5 are displayed $\Pi$ (eq. \ref{pi}) 
and $M$ (eq. \ref{m}) for the observations and
the $\Omega=0.4$ model.  Symbols are as in previous figures, and no errors
are included in the model of figure 5a, while in figure 5b we show the
results with TF scatter and zero-point offsets.
Both affect significantly the coherence of the velocities;  
TF zero-point offsets rise $\Pi$ in a factor proportional to $\Delta r$; while
the scatter reduces the coherence at large separations $\Delta r$
and increases it at small $\Delta r$.
It is interesting to compare these results with those derived from
the velocity correlation function (see for instance \cite{strauss})
where a coherence length of $2000$ $km/s$
is observed for the Aaronson sample of the MarkIII catalog. From our analysis
we find a larger value of the coherence length $\simeq 2800$ $km/s$. 
However we note that this spatial scale defined by a lack of velocity
correlation depends on the catalog depth, and very strongly
on TF scatter since at distances $\geq 2000$ $km/s$, 
the associated uncertainties in the 
peculiar velocities are of comparable magnitude to the flow coherence.

\begin{figure}[bt]
\centerline{\epsfxsize=\textwidth \epsffile{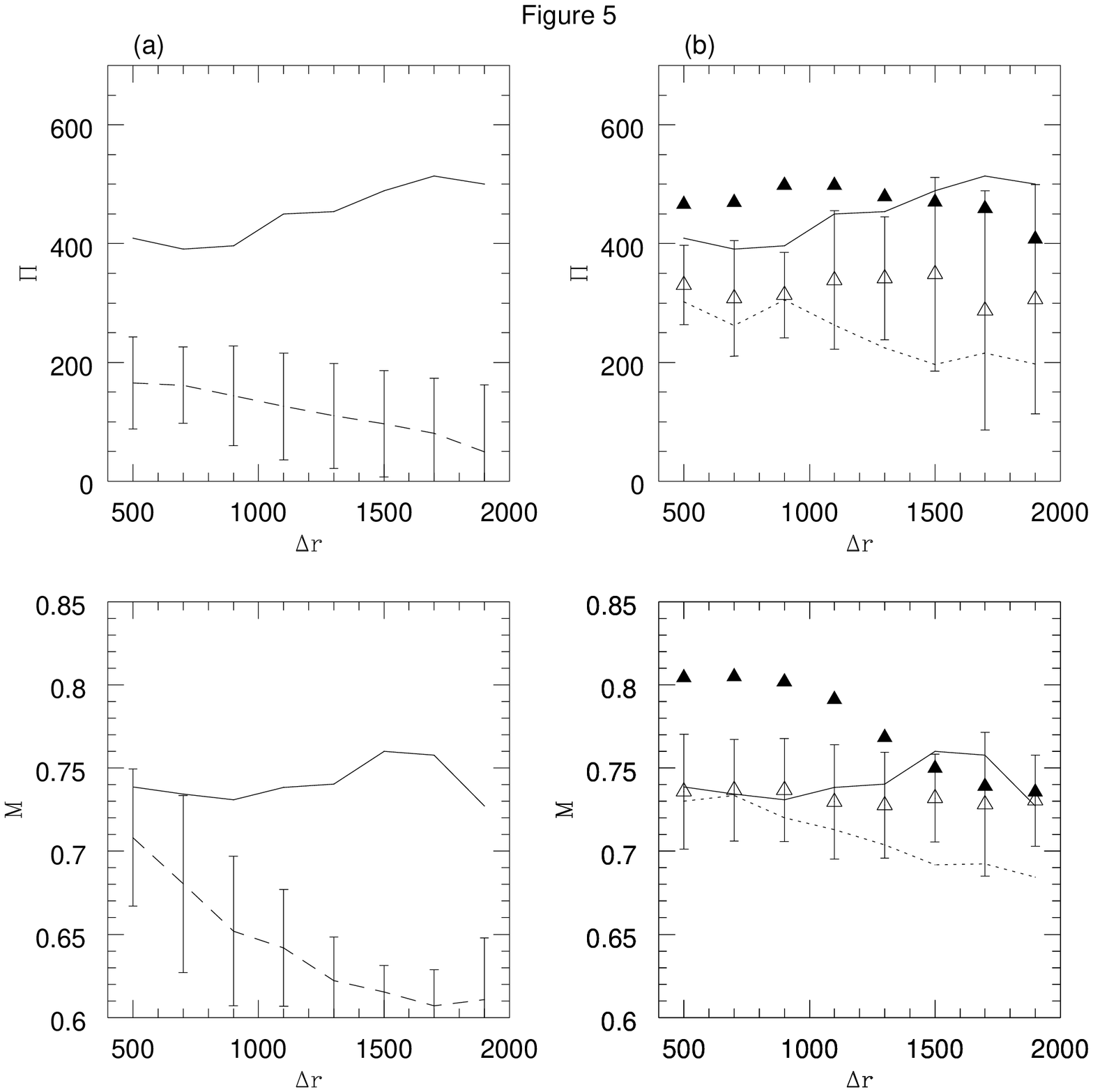}}
\caption{
$\Pi$ and $M$ test results.  Solid lines show the observed values.
Figure 5a: no errors included in the $\Omega=0.4$ (dashed line).
Figure 5b: effects of a $0.35$ $mag$  TF scatter (dotted line) and additional
$+0.15$ $mag$ and $-0.15$ $mag$ TF zero-point offsets
(open and filled triangles) in the $\Omega=0.4$ CDM model. 
Error bars represent the same as in previous figures.
}
\label{fig:f5}
\end{figure}

In order to asses the ability of the models to reproduce the observations
we calculate the frequency $\cal L$$(\Omega)$ that the observational 
values of a statistical test are within the range 
of the results obtained with a large number of random observers (mock
catalogs) as a
function of model density parameter $\Omega$.  Specifically, we compute

$$\chi_{obs}^2=\sum_{\Delta r}(O_{\Delta r}-A_{\Delta r})^2
/\sigma_{\Delta r}^2$$

$$\chi_{i}^2=\sum_{\Delta r}(R_{\Delta ri}- 
A_{\Delta r})^2/\sigma_{\Delta r}^2$$

\noindent {where R refers to the 
resulting value of a particular statistical test
for the random $i^{th}$ observer; and A, to the corresponding 
average across the ensemble of random observers.
Both R and A are computed for a given model as a 
function of separation $\Delta r$, 
$\sigma_{\Delta r}^2=\sum_i (R_{\Delta ri}-A_{\Delta r})^2/N$, 
 and O corresponds to the observed value of the statistical tests under
consideration.
}

We find that $\chi_i^2$ has approximately a gaussian distribution with 
dispersion $\sigma_{i}$ in the models explored, and we therefore estimate 
\begin{equation}
{\cal L}(\Omega)= \frac{1}{\sigma_i} \sqrt{\frac{2}{\pi}} 
\int^{\infty}_{\chi_{obs}^2}
e^{-\chi^{4}/(2 \pi \sigma_{i}^{2})} 
 \,d\chi^2
\end{equation}
as a suitable likelihood of a given model to reproduce the observed
values of a particular statistical test. 

For the two catalog depths analyzed $d_{lim}=5000$ $km/s$ and 
$d_{lim}=10000$ $km/s$ we find that the results of the statistics do not 
change strongly.  Moreover, the changes in the statistics 
from one catalog depth to the other in the observations and in the mock catalogs
are similar indicating that the conclusions derived are not depth dependent.
However, the dispersion arising from different observer 
positions is considerably larger in the $d_{lim}=5000 $ $km/s$ case 
which suggests the use of the deeper sample.

We display in figure 6 the likelihood ${\cal L}(\Omega)$ obtained for 
our different statistical tests 
using catalog depth $d_{lim}=10000$ $km/s$, and considering
a $0.35$ $mag$ TF scatter (figure 6a), plus
additional $-0.15$ and $+0.15$ $mag$ zero-point offsets (figure 6b and c).
>From inspection to figure 6a, we note that in the absence of a
TF zero-point offset 
the likelihood associated to $D$ and $\Delta V$ for models 
with $\Omega>0.6$ are $\leq 0.1\%$.
On the other hand, the observed flow coherence as reflected in 
the $\Pi$ and $M$ tests is in better agreement with high $\Omega$ models.
The effects of a negative and positive TF zero-point offset may be 
appreciated in figure 6b and 6c, where it is seen that  
high values of $\Omega>0.6$ are not suitable to
match the observed $D$, $D1$, and $\Delta V$ statistical 
tests, while the remaining tests 
do not exclude any of the models explored (${\cal L} > 10 \%$).  
We find that a positive TF zero-point offset provides a better 
agreement between the observations and low density CDM models. 

\begin{figure}[bt]
\centerline{\epsfxsize=\textwidth \epsffile{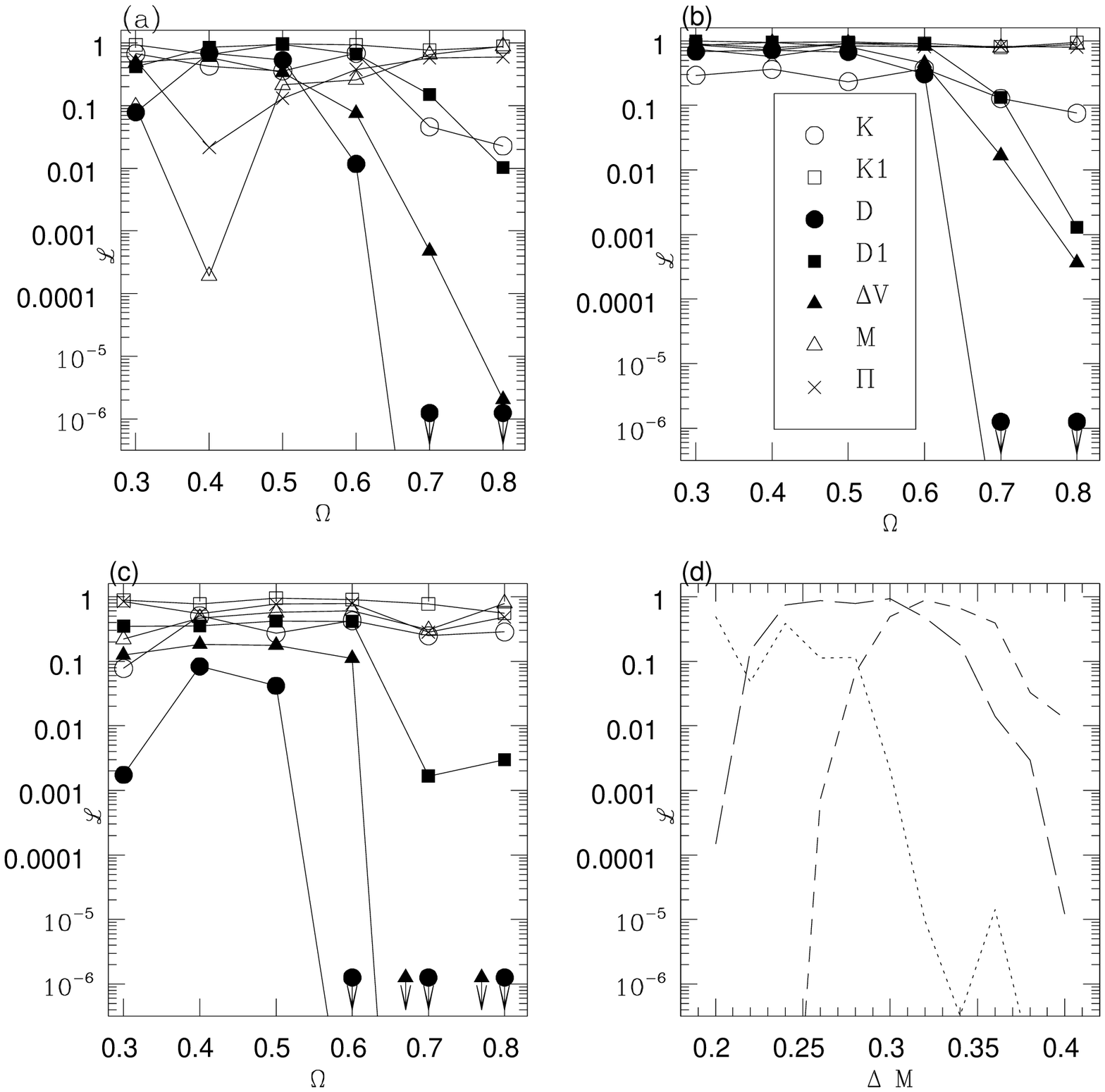}}
\caption{
Likelihood estimates of the models for the different statistical tests
(eq. 9) as a function of density parameter
$\Omega$ (figures 6a to 6c). 
The adopted catalog depth is $d_{lim}=10000$ $km/s$.
Open circles: K, open squares: K1, filled circles: D, filled squares: D1, 
filled triangles: $\Delta V$, crosses: $\Pi$, and open triangles: $M$.
Figure 6a: TF scatter $\Delta M=0.35$ $mag$ considered in the mock catalogs.
Figure 6b: same as figure 6a with an additional $-0.15$ $mag$ zero-point 
offset.
Figure 6c: same as figure 6a with an additional $+0.15$ $mag$ zero-point 
offset.
In figure 6d we show the Likelihood estimates for the open CDM models 
$\Omega=0.4$, $\Omega=0.6$ and $\Omega=0.8$
(dashed, long dashed and dotted lines respectively),
for the $\Delta V$ test as a function of the magnitude of the TF scatter 
$\Delta M$.
}
\label{fig:f6}
\end{figure}

We have also analyzed the 
dependence of the results on the
magnitude of the TF scatter.  In figure 6d, we display $\cal L$ for the
$\Delta V$ test for the $\Omega=0.4$, $\Omega=0.6$ and $\Omega=0.8$ 
CDM models as a function of the TF scatter $\Delta M$.  
It can be seen in the figure
the strong dependence of $\cal L$ on the assumed 
value of the TF scatter.  For $\Delta M>0.3$ the $\Omega=0.8$ model is ruled
out by this test with large confidence.  On the other hand if $\Delta M<0.25$
the $\Omega =0.4$ model would be ruled out.  However it should be noticed that
the TF scatter of the different samples in the MarkIII catalog, are 
in all cases $>0.35$ (See Table 1), the adopted value in our analysis.

We have repeated our analysis using inverse TF distances to test
possible systematic influences on the statistics.  
We find similar results in this case than in the case of forward distances
indicating that either estimates are suitable to be used 
in our statistical studies. 

We have also examined the effects of biasing 
the distribution of particles in the 
numerical simulations on the results of the statistics. 
We have biased the $\Omega=0.4$ model and
anti-biased the $\Omega=0.8$ model, so that the relative galaxy number 
fluctuation on spheres of $8$ Mpc is unity for these two models. 
Each particle is associated to a galaxy with probability $\eta$ which depends 
on the local density.
We smooth the density field calculating
the mass density $\rho$ centered at each particle within a sphere of radius
$1.5 h^{-1}$  $Mpc$ and we adopt a simple power-law model 
$\eta(\rho) = (\rho /\rho_{c})^{\alpha}$ for the assignment of particles.  
We find that our results 
are not strongly modified for these biased models with resulting
${\cal L}(\Omega)$ of the same order of magnitude. 
This result is important since it indicates that our conclusions regarding the 
value of the density parameter do not depend crucially on the adopted 
normalization nor on the assumption that galaxies trace the mass.

\section{Conclusions}

In this work we have attempted to asses the effects of  
galaxy distance uncertainties
and biases on statistical tests of the peculiar velocity field 
using samples of spiral galaxies from the Mark III compilation
and mock catalogs taken from numerical simulations.
Our studies comprise relative
velocity tests and pair velocity correlations as a function of 
separation, as well as a bulk flow analysis determined on spheres of $10 h^{-1}$
$Mpc$ radius. 
We construct mock catalogs using numerical 
simulations corresponding to COBE normalized CDM models with 
different values of density parameter and cosmological constant. 
The models take into account the Tully-Fisher (TF) 
scatter and possible TF zero-point
shift, as well as variations in the results of the simulations
arising from different observer positions.
The analysis of the departures from gaussianity of the observational
velocity distributions
as measured by the kurtosis tests $K$ and $K1$ show
 that only a small fraction ($<2 \%)$ of possible 
spuriously high velocities (and therefore biased distance estimates)
 might be present in the observational data.
We find that the observed scatter of the TF relation
plays a very important
role when deriving constraints to the cosmological parameters.
If errors in galaxy distance estimates are neglected, the observed magnitude of
the $\Delta V$ results show consistency with high 
density CDM models ($\Omega>0.8$). 
When the observed TF scatter $\Delta M \simeq 0.35$ is taken into account
a significant disagreement with observations is found for the $\Delta V$ 
and $D1$ statistics in models with $\Omega > 0.6$.

A possible offset in the TF zero-point used to determine
the distances of the galaxies in the observational data artificially enhance 
velocity correlations ($\Pi$ and $M$ statistics). 
Due to this fact, if zero-point offsets of 
the order of a tenth of a magnitude are considered,  
the  CDM models explored provide a more
satisfactory fit to observations in these tests. 

The results presented in this work are not sensitive to sample variations.
We have applied the same statistical tests to the subsample corresponding to
Mathewson data and we find
similar results than those of the total sample.
It should be noted that the variations on the statistics arising 
from different observers are significantly enhanced when errors on
the peculiar velocities are included.  This fact make more difficult
the distinction between the models. 
Error bars in figures 1 to 5 show the variations of the results
arising from different observer positions in the models and serve as a test
for the dependence of the observational results on our particular position
in space. 
In our analysis of biased models we have not found relevant 
differences in the results of the statistical 
analysis of samples with different density thresholds.  
The selection of low (high)
density particles in the simulations does not produce very relevant 
effects in the statistics although lower (higher)
values of $\Pi$ and $\Delta V$ are observed due to the 
oversampling (undersampling) of high density particles where random
motions dominate.
Finally when comparing models and observations, we find that neither the
K, K1, M, nor $\Pi$ statistical tests are sensitive to the density 
parameter of the models.

We find crucial to consider properly the intrinsic scatter of the Tully-Fisher 
relation in studies of the peculiar velocity field. 
Although low values of the density parameter $\Omega < 0.6 $, are favored 
by our statistical
analysis  $D1$ and $\Delta V$ when this scatter is taken into account 
$\pm 0.15$ $mag$ zero-point offsets in the TF relation 
are required in order to provide 
the observed values of $\Pi$ and $M$ statistics in the models explored.
A positive TF zero-point offset of this magnitude 
would imply lower values of the Hubble constant which may be worth 
to consider in a controversial  topic (see for instance \cite{shanks} and 
\cite{th} and references therein).

\section{Acknowledgements}

We thank H. Couchmann for kindly providing the AP3M code.
We have benefited from helpful discussions with Jim Peebles and
Julio Navarro. 

This work was supported by the Consejo de Investigaciones Cient\'\i ficas y
T\'ecnicas de la Rep\'ublica Argentina, CONICET, the Consejo de
Investigaciones Cient\'\i ficas y Tecnol\'ogicas de la Provincia de C\'ordoba,
CONICOR, and the Fundaci\'on Antorchas, Argentina.
{}

\end{document}